\documentclass[11pt,a4paper]{article}

\usepackage{epsfig,amsmath,amssymb,cite}

\begin{document}

\title{Gravitational lensing by massless braneworld black holes}
\author{Ernesto F. Eiroa$^{1,2,}$\thanks{e-mail: eiroa@iafe.uba.ar}, Carlos M. Sendra$^{1}$\thanks{e-mail: cmsendra@iafe.uba.ar} \\
{\small $^1$ Instituto de Astronom\'{\i}a y F\'{\i}sica del Espacio, C.C. 67, Suc. 28, 1428, Buenos Aires, Argentina}\\
{\small $^2$ Departamento de F\'{\i}sica, Facultad de Ciencias Exactas y 
Naturales,} \\ 
{\small Universidad de Buenos Aires, Ciudad Universitaria Pab. I, 1428, 
Buenos Aires, Argentina} }
\maketitle
\date{}

\begin{abstract}
In this paper, we study massless braneworld black holes as gravitational lenses. We find the weak and the strong deflection limits for the deflection angle, from which we calculate the positions and magnifications of the images. We compare the results obtained here with those corresponding to Schwarzschild and Reissner--Nordstr\"om spacetimes, and also with those found in previous works for some other braneworld black holes. 
\end{abstract}



\maketitle

\section{Introduction}

In braneworld cosmological models, the ordinary matter is on a three dimensional space called the brane, which is embedded in a larger space denominated the bulk, where only gravity can propagate. These models were proposed to solve the hierarchy problem, i.e. the difficulty in explaining why the gravity scale is sixteen orders of magnitude greater than the electro-weak scale. Motivated by string theory (M-theory), they have received great attention in recent years \cite{bwrev}. The properties of black holes will be different due to the presence of the extra dimensions \cite{kanti}. In the simplest of the braneworld scenarios, the Randall-Sundrum \cite{rsII} model (a positive tension brane in a bulk with one extra dimension and a negative cosmological constant), primordial black holes formed in the high energy epoch would have a longer lifetime \cite{cgl}, because of a different evaporation law. These primordial black holes could have a growth of their mass through accretion of surrounding radiation during 
the high energy phase, increasing their lifetime \cite{majumdar}, so they might have survived up to the present. Within these cosmological models, high energy collisions in particle accelerators or in cosmic rays could also create black holes \cite{kanti}. In the Randall-Sundrum scenario, a spherically symmetric black hole solution on a three dimensional brane was found \cite{dmpr}, characterized by a tidal charge due to gravitational effects coming from the fifth dimension. A general class of braneworld black holes with spherical symmetry was presented in Ref. \cite{bronnikov}. Within this class, there are black hole solutions without mass, i.e. where the curvature is produced only by a tidal effect \cite{bronnikov}.

The discovery of supermassive black holes at the center of galaxies, including our own, has lead to a growing interest in the study of black holes as gravitational lenses. The observable quantities, such as the positions, magnifications, and time delays of the relativistic images, produced by photons passing close to the photon sphere, can be calculated by using the strong deflection limit, which was introduced by Darwin \cite{darwin} for the Schwarzschild spacetime. This method, consisting in a logarithmic approximation of the deflection angle, was rediscovered by other authors \cite{otros}, and then extended to the Reissner--Nordstr\"om metric \cite{eiroto}, and to any spherically symmetric object with a photon sphere \cite{boz}. Numerical studies of black hole lenses were also done \cite{numerical}. Other interesting works considering strong deflection lenses with spherical symmetry can be found in Refs. \cite{alternative,nakedsing1,nakedsing2}. In particular, non-rotating braneworld black holes were 
analyzed as lenses \cite{bwlens1,bwlens2,bwlens3,bwlens4} in recent years. The optical effects of Kerr black holes were analyzed by several authors \cite{bozza1,bozza2,vazquez,kraniotis}. The apparent shapes or shadows of rotating black holes present an optical deformation due to the spin, topic which has been recently examined in several articles \cite{falcke,shadow,bozza2,zakharov}, in the belief that direct observation of these objects will be possible in the near future \cite{zakharov}. Optical properties of rotating braneworld black holes \cite{aliev} were studied in Refs. \cite{shbw}. A recent review of black hole lensing, with a discussion of the observational prospects, can be found in Ref. \cite{bozzareview}.

In this article, we study massless black holes as gravitational lenses, in the Randall-Sundrum braneworld scenario. In Sec. 2, we introduce the metric and we find the exact expression of the deflection angle. In Sec. 3, we perform the weak deflection limit and we obtain the positions and magnifications of the primary and secondary images. In Sec. 4, we find the strong deflection limit, from which we calculate the positions and magnifications of the relativistic images. Finally, in Sec. 5, we present a discussion of the results. We adopt units such that $G=c=1$.

\section{Deflection angle}

We start from the spherically symmetric geometry \cite{bronnikov}:
\begin{equation}
ds^{2}=-A(r)dt^{2}+B(r)dr^{2}+C(r)(d\theta^{2}+\sin^{2}\theta d\phi^{2}),
\label{m1}
\end{equation}
where the metric functions are given by
\begin{equation}
A(r)=1-\frac{h^2}{r^2},  \hspace{0.5cm} B(r)^{-1}=\left( 1-\frac{h^2}{r^2} \right) \left( 1+\frac{\kappa -h}{\sqrt{2r^2-h^2}} \right), \hspace{0.5cm} C(r)=r^2,
\label{m2}
\end{equation}
with $\kappa $ and $h>0$ constants. It is useful to adimensionalize all quantities with $h$,  by introducing $x=r/h$, $T=t/h$, and $\eta =\kappa /h$, so the metric takes the form
\begin{equation}
ds^{2}=-A(x)dT^{2}+B(x)dx^{2}+C(x)(d\theta^{2}+\sin^{2}\theta d\phi^{2}),
\label{m3}
\end{equation}
where
\begin{equation}
A(x)=1-\frac{1}{x^2},  \hspace{0.5cm} B(x)^{-1}=\left( 1-\frac{1}{x^2} \right) \left( 1+\frac{\eta -1}{\sqrt{2x^2-1}} \right), \hspace{0.5cm} C(x)=x^2.
\label{m4}
\end{equation}
If $\eta >0$ the geometry corresponds to a black hole \cite{bronnikov} with a simple horizon at the surface $x_h=1$, while if $\eta =0$ this horizon is double. When $\eta <0 $, there is a symmetric wormhole throat outside the horizon \cite{bronnikov} with radius $x_{th}=\sqrt{1+(1-\eta )^2}/ \sqrt{2}>1$, where $B(x)^{-1}$ has a simple zero. We are solely interested in black holes, so we adopt $\eta >0$. These braneworld black holes have no matter and no mass, and they only exist as a tidal effect of the bulk gravity  \cite{bronnikov}. The effective energy-momentum tensor on the brane (see \cite{bronnikov}) does not satisfy the null energy condition (then the weak and strong energy conditions are also violated) in the region outside the horizon, i.e. it is exotic. The simplest metric corresponds to $\eta =1$, leading to $A(x)=B(x)^{-1}=1-1/x^2$, which can be identified as the Reissner--Nordstr\"om metric with zero mass and a purely imaginary charge, having the horizon at $x_h=1$ and the singularity at $x=0$. 
When $0\le \eta<1$, the horizon is at $x_h=1$, and the geometry have a point singularity at $x=0$ and a singular surface at $x=1/\sqrt{2}$; between this surface and the horizon there is a throat with radius $1/\sqrt{2}<x_{th}=\sqrt{1+(1-\eta )^2}/ \sqrt{2}\le 
1$; note that the throat is covered by the horizon, so we speak of a black hole instead of a wormhole. In the case with $\eta >1$, we have again the horizon at $x_h=1$, a point singularity at $x=0$, and a singular surface at $x=1/\sqrt{2}$, but the throat is not present because the function $B(x)^{-1}$ has no zeros.
 
\begin{figure}[t!]
\begin{center}
    \includegraphics[scale=0.70,clip=true,angle=0]{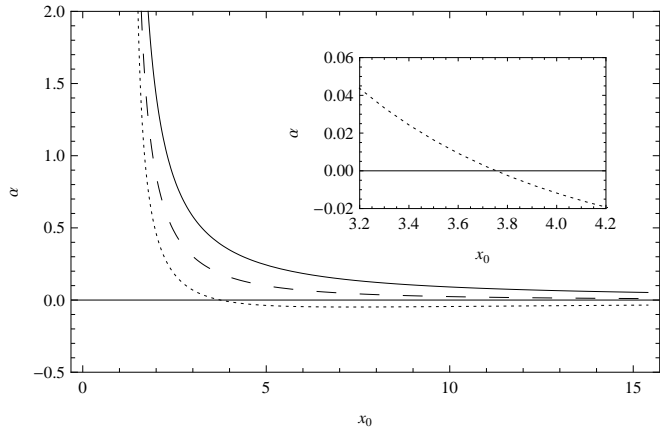}
    \caption{The deflection angle $\alpha $ as a function of the (dimensionless) closest approach distance $x_0=r_0/h$, for the cases $0<\eta <1$ (shown $\eta=0.1$, solid line), $\eta=1$ (dashed line) and $\eta >1$ (shown $\eta=2$, dotted line). Note that for $\eta >1$ the deflection angle becomes negative as $x_0$ grows.}
    \label{alfaexactofig}
\end{center}
\end{figure}

The deflection angle for a photon coming from infinity can be written as a function of the adimensionalized closest approach distance $x_0=r_0/h$, in the form \cite{weinberg,nakedsing1}
\begin{equation}
\alpha(x_0)=I(x_0)-\pi, 
\label{alfa1}
\end{equation}
where 
\begin{equation}
I(x_0)=\int^{\infty}_{x_0}\frac{2\sqrt{B(x)}dx}{\sqrt{C(x)}\sqrt{A(x_0)C(x)\left[ A(x)C(x_0)\right]^{-1}-1}}. 
\label{i0}
\end{equation}
The deflection angle $\alpha $ grows as $x_0$ approaches to the photon sphere radius $x_{ps}$, where diverges. Replacing the metric functions (\ref{m2}) in Eq. (\ref{i0}), the exact deflection angle is obtained, and  it is plotted in Fig. \ref{alfaexactofig} for different values of $\eta$. In the case $\eta>1$ the deflection angle becomes negative from a certain value (larger than $x_{ps}$) to infinity.

\section{Weak deflection limit}

When $x_0\gg x_{ps}$, the deflection angle is small. Defining the variables $k=x_0/x$ and $z=1/x_0$, the integral (\ref{i0}) has the form
\begin{equation}
I(x_0)=2\int^{1}_{0} \frac{1}{\sqrt{1-k^2}}h(k)dk,
\label{i1}
\end{equation}
where
\begin{equation}
h(k)=\frac{1}{\sqrt{\left( -z^2 - k^2 z^2 + 1\right)  \left[1 + (-1 + \eta) z \left( \sqrt{-z^2 + 2k^{-2}}\right) ^{-1}\right]}}.
\label{i2}
\end{equation}
Expanding $h(k)$ to first order in $z=0$, replacing it in Eq. (\ref{i1}), integrating all the terms separately and rewriting $x_0=1/z$, the deflection angle as a function of the closest approach distance, in the weak deflection limit, finally adopts the form
\begin{equation}
\alpha(x_0)\approx\frac{(1-\eta)}{\sqrt{2}}\frac{1}{x_0}.
\label{alfawd1}
\end{equation}
It is convenient to express this angle in terms of the dimensionless impact parameter $u$, which is related to the closest approach distance $x_0$ by \cite{weinberg,nakedsing1}
\begin{equation}
u=\sqrt{\frac{C(x_0)}{A(x_0)}};
\label{impactu}
\end{equation}
so inverting it we have
\begin{equation}
x_0=\sqrt{\frac{u^2+\sqrt{u^2\left(u^2-4\right)}}{2}}.
\label{impact}
\end{equation}
Replacing Eq. (\ref{impact}) in Eq. (\ref{alfawd1}) and making a first order expansion in $1/u$, for large $u$ the deflection angle becomes
\begin{equation}
\alpha(u)\approx\frac{(1-\eta)}{\sqrt{2}}\frac{1}{u}.
\label{alfawd2}
\end{equation}
The optical axis can be defined as the line joining the observer ($o$) and the lens ($l$). The angular positions of the source ($s$) and the images, seen from the observer, are $\beta $ (taken positive) and $\theta $, respectively. From the lens geometry it is clear that $u=d_{ol}\sin\theta\approx d_{ol}\theta$, with $d_{ol}$ the adimensionalized observer-lens angular diameter distance; so the deflection angle in the weak deflection limit takes the form
\begin{equation}
\alpha(\theta)\approx\frac{(1-\eta)}{\sqrt{2}d_{ol}}\frac{1}{\theta}.
\label{alfawd3}
\end{equation}
When $0<\eta<1$, the leading term of the deflection angle is positive and for $\eta>1$ this term becomes negative, which is in accordance with Fig. \ref{alfaexactofig}. For $\eta=1$, the first order term vanishes and a higher order expansion is required. 

For a small deflection angle, i.e. light rays passing far from the photon sphere, the lens equation is given by the expression
\begin{equation}
\beta=\theta-\frac{d_{ls}}{d_{os}}\alpha ,
\label{lenseq}
\end{equation}
where $d_{ls}$ and $d_{os}$ are the adimensionalized lens-source and observer-source angular diameter distances, respectively. For perfect alignment of the source, lens and observer ($\beta=0)$, the Einstein ring is formed, and for $0<\eta<1$ its radius is given by
\begin{equation}
\theta_{E}=\sqrt{\frac{(1-\eta)d_{ls}}{\sqrt{2}d_{ol}d_{os}}}.
\label{einstein1}
\end{equation}
In terms of this quantity and conserving only the $1/\theta $ term in Eq. (\ref{alfawd3}), when $0<\eta <1$ the deflection angle becomes
\begin{equation}
\alpha(\theta)\approx\frac{\theta^{2}_{E}d_{os}}{d_{ls}}\frac{1}{\theta}.
\label{alfaw1}
\end{equation}
Replacing (\ref{alfaw1}) in the lens equation (\ref{lenseq}), two solutions for the angular positions of images are found:
\begin{equation}
\theta_{+}=\frac{\beta+\sqrt{\beta^2+4\theta^{2}_{E}}}{2}
\label{thetamas}
\end{equation}
and
\begin{equation}
\theta_{-}=\frac{\beta-\sqrt{\beta^2+4\theta^{2}_{E}}}{2},
\label{thetamenos}
\end{equation}
corresponding to the primary and the secondary images. 

If $\eta=1$, from Eq. (\ref{alfawd3}) we see that the first order term vanishes in the expansion of the deflection angle, so the approximation adopted here (first order in $1/u$) is no longer valid. So, for $\eta =1$, higher order or numerical methods are required to obtain the positions of the two weak deflection images. When $\eta >1$, the deflection angle is small and negative for large values of  $x_0$, as it can be seen from Fig. \ref{alfaexactofig}. In this case, the black hole then behaves like a divergent lens. For small $\beta $, the lens equation (\ref{lenseq}) has no real solutions, which means that the deflected photons never reach the observer and no weak deflection images are formed. Also, for $\beta =0$ the Einstein ring is not present for large $x_0$. As shown in Fig. \ref{alfaexactofig}, there is a value of $x_{0}$ near to the photon sphere but not too close to it, namely  $x_z$, for which the deflection angle is zero. For any $x_0$ larger than $x_z$, we have that $\alpha (x_0)<0$ and a similar reasoning as above leads to no images for high alignment, and no Einstein ring if $\beta =0$. But when $x_0<x_z$ the deflection angle is positive and the lens is convergent. For values of $x_0$ slightly smaller than $x_z$, the small and positive deflection angle gives way to two images for high alignment and an Einstein ring if $\beta =0$. In this case, the approximation given by Eq. (\ref{alfawd1}) no longer holds, and one has to rely on numerical methods for obtaining the positions of the images, which is outside the scope of our work.

\begin{figure}[t!]
\begin{center}
    \includegraphics[scale=0.70,clip=true,angle=0]{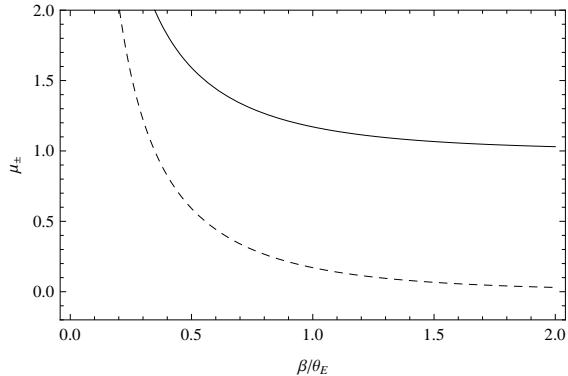}
    \caption{The magnifications of the primary (solid line) and secondary (dashed line) images for $0<\eta<1$, as functions of the quotient of the angular position of the source $\beta$ and the angular Einstein radius $\theta_E $, which is a function of $\eta $ and the lensing distances (see text). }
    \label{magnificationfig}
\end{center}
\end{figure}

The quotient of the solid angles subtended by the image and the source gives the magnification:
\begin{equation}
\mu =\left|\frac{\sin\beta}{\sin\theta}\frac{d\beta}{d\theta}\right|^{-1},
\label{magnif1}
\end{equation}
which for small angles reduces to
\begin{equation}
\mu =\left|\frac{\beta}{\theta}\frac{d\beta}{d\theta}\right|^{-1}.
\label{magnif2}
\end{equation}
For $0<\eta<1$, and using equations (\ref{thetamas}) and (\ref{thetamenos}), the magnifications of the primary and secondary images are given by
\begin{equation}
\mu_{+}=\frac{\left(\beta+\sqrt{\beta^2+4\theta^{2}_{E}}\right)^2}{4\beta\sqrt{\beta^2+4\theta^{2}_{E}}},
\label{mag1}
\end{equation}
and
\begin{equation}
\mu_{-}=\frac{\left(\beta-\sqrt{\beta^2+4\theta^{2}_{E}}\right)^2}{4\beta\sqrt{\beta^2+4\theta^{2}_{E}}}.
\label{mag2}
\end{equation}
The magnifications of the weak deflection images are plotted in Fig. \ref{magnificationfig}, as functions of the angular position of the source divided by the corresponding Einstein radius, for the case $0<\eta<1$. Note that the magnifications increase as $\beta $ decreases, i.e. when the alignment is higher.

\section{Strong deflection limit}

\begin{figure}[t!]
\begin{center}
    \includegraphics[scale=0.90,clip=true,angle=0]{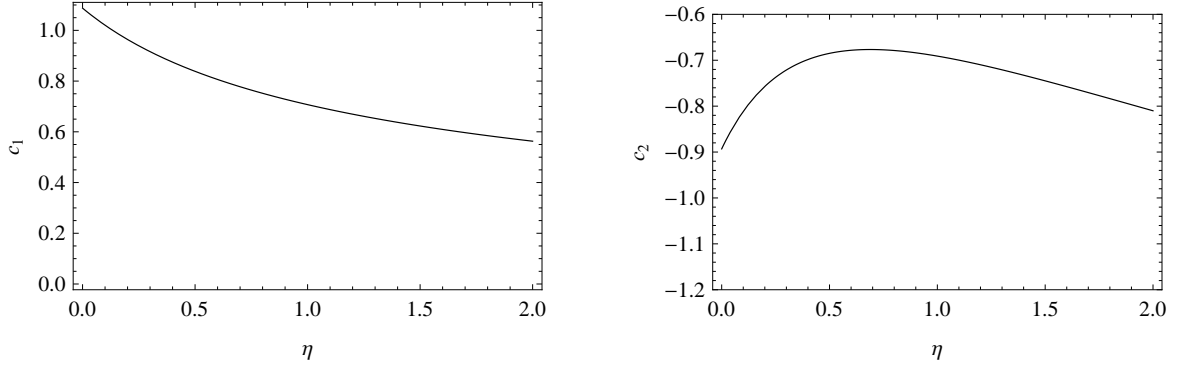}
    \caption{Strong deflection limit coefficients $c_1$ and $c_2$ as functions of the parameter $\eta$.}
    \label{coef}
\end{center}
\end{figure}

Now, we consider the case of the images produced by photons passing close to the photon sphere. The radius $x_{ps}$ of the photon sphere is given by the largest positive solution of the equation
\begin{equation}
\frac{A'(x)}{A(x)}=\frac{C'(x)}{C(x)},
\label{xps}
\end{equation}
where the prime represents the derivative with respect to $x$. For the massless black hole is the constant value $x_{ps}=\sqrt{2}$. We take the observer-source $d_{os}$, observer-lens $d_{ol}$ and the lens-source $d_{ls}$ angular diameter distances much greater than the horizon radius $x_h$. In this approximation, the lens equation has the form \cite{bozzale}
\begin{equation}
\tan \beta =\frac{d_{ol}\sin \theta - d_{ls} \sin (\alpha -\theta)}{d_{os} \cos (\alpha -\theta)} .
\label{pm1}
\end{equation}
The lensing effects are more important when the objects are highly aligned, so we will only study this case, in which the angles $\beta $ and $\theta $ are small, and $\alpha $ is close to an even multiple of $\pi $. When $\beta \neq 0$ two infinite sets of point relativistic images are obtained. The first set of relativistic images have a deflection angle that can be written as $\alpha =2n\pi +\Delta \alpha _{n}$, with $n\in \mathbb{N}$ and $0<\Delta \alpha _{n}\ll 1$. In this approximation, the lens equation results \cite{boz,bozzale}
\begin{equation}
\beta =\theta -\frac{d_{ls}}{d_{os}}\Delta \alpha _{n}.
\label{pm2}
\end{equation}
For the other set of images, $\alpha =-2n\pi -\Delta \alpha _{n}$, then $\Delta \alpha _{n}$ is replaced by $-\Delta \alpha _{n}$ in Eq. (\ref{pm2}). For the calculation of the deflection angle for photons passing close to the photon sphere, following Ref. \cite{boz}, it is useful to separate the integral as a sum
\begin{equation}
I(x_0)=I_D(x_0)+I_R(x_0), 
\label{i0n}
\end{equation}
of a divergent part
\begin{equation}
I_D(x_0)=\int^{1}_{0}R(0,x_{ps})f_0(z,x_0)dz 
\label{id}
\end{equation}
and a regular part
\begin{equation}
I_R(x_0)=\int^{1}_{0}[R(z,x_0)f(z,x_0)-R(0,x_{ps})f_0(z,x_0)]dz, 
\label{ir}
\end{equation}
where
\begin{equation}
z=\frac{A(x)-A(x_0)}{1-A(x_0)},
\label{z}
\end{equation}
\begin{equation}
R(z,x_0)=\frac{2\sqrt{A(x)B(x)}}{A'(x)C(x)}[1-A(x_0)]\sqrt{C(x_0)},
\label{r}
\end{equation}
\begin{equation}
f(z,x_0)=\frac{1}{\sqrt{A(x_0)-[(1-A(x_0))z+A(x_0)]C(x_0)[C(x)]^{-1}}},
\label{f}
\end{equation}
and
\begin{equation}
f_0(z,x_0)=\frac{1}{\sqrt{\varphi (x_0) z+\gamma (x_0) z^{2}}}, 
\label{f0}
\end{equation}
with
\begin{equation}
\varphi (x_0)=\frac{1-A(x_0)}{A'(x_0) C(x_0)}\left[ A(x_0) C'(x_0) - A'(x_0) C(x_0)\right],
\label{varphi}
\end{equation}
and
\begin{eqnarray}
\gamma (x_0) &=& \frac{\left[ 1-A(x_0)\right] ^{2}}{2[A'(x_0)]^{3} [C(x_0)]^{2}}\left\{ 2 [A'(x_0)]^{2} C(x_0) C'(x_0) - A(x_0) A''(x_0) C(x_0) C'(x_0) \right. \nonumber \\
&& \left. + A(x_0) A'(x_0) \left[ C(x_0) C''(x_0) -2 [C'(x_0)]^{2}\right] \right\} .
\label{gamma}
\end{eqnarray}
When $x_0 \neq x_{ps}$, it is easy to see that $\varphi \neq 0$ and $f_0\sim 1/\sqrt{z}$, then the integral $I_D(x_0)$ converges. But if $x_0= x_{ps}$, by using Eq. (\ref{varphi}) we obtain that $\varphi =0$ and $f_0\sim 1/z$, thus $I_D(x_0)$ has a logarithmic divergence. Then, $I_D(x_0)$ is the term which contains the divergence at $x_0=x_{ps}$ and $I_R(x_0)$ is regular because the divergence has been subtracted. The impact parameter $u$ is more easily connected to the lensing parameters than $x_0$; they are related by Eq. (\ref{impactu}). The logarithmic divergence of the deflection angle for photons passing close to the photon sphere, in terms of the impact parameter, has the general form \cite{boz}
\begin{equation}
\alpha(u)=-c_1\ln\left(\frac{u}{u_{ps}}-1\right)+c_2+O(u-u_{ps}),
\label{alfasdl}
\end{equation}
where $u_{ps}$ is the impact parameter evaluated at $x_0=x_{ps}$,
\begin{equation}
c_1=\frac{R(0,x_{ps})}{2\sqrt{\gamma(x_{ps})}}
\label{c1}
\end{equation}
and
\begin{equation}
c_2=-\pi+c_R+c_1\ln \frac{2\gamma(x_{ps})}{A(x_{ps})},
\label{c2}
\end{equation}
with
\begin{equation}
c_R=I_R(x_{ps}).
\label{cr}
\end{equation}
This logarithmic approximation is called the strong deflection limit. For the massless braneworld black hole, using that $x_{ps}=\sqrt{2}$, we obtain that $u_{ps}=2$, $A(x_{ps})=1/2$, $\gamma (x_{ps})=1/2$, and $R(0,x_{ps})=\sqrt{\sqrt{3}/(\sqrt{3}-1+\eta )}$. Then, from Eqs. (\ref{c1}) and (\ref{c2}), we find that the coefficients of the strong deflection limit are given by
\begin{equation}
c_1=\sqrt{\frac{\sqrt{3}}{2(\sqrt{3}-1+\eta)}}
\label{c1bh}
\end{equation}
and
\begin{equation}
c_2=-\pi+c_R+c_1\ln2,
\label{c2bh}
\end{equation}
where $c_R$ have to be calculated numerically for each value of $\eta$, because the integral (\ref{ir}) cannot be obtained analytically. The strong deflection limit coefficients $c_1$ and $c_2$ are plotted in Fig. \ref{coef}.

The deflection angle given by Eq. (\ref{alfasdl}) can be written in terms of the image position $\theta$, since $u=d_{ol}\sin\theta\approx d_{ol}\theta$: 
\begin{equation}
\alpha (\theta )\approx-c_{1}\ln \left( \frac{d_{ol}\theta }{u_{ps}}-1 \right) 
+c_{2}. 
\label{pm4}
\end{equation}
By inverting Eq. (\ref{pm4}) and performing a first order Taylor expansion around $\alpha=2n\pi$, the angular position of the $n$-th image is obtained:
\begin{equation}
\theta _{n}=\theta ^{0}_{n}-\zeta _{n}\Delta \alpha _{n},
\label{pm6}
\end{equation}
where
\begin{equation}
\theta ^{0}_{n}=\frac{u_{ps}}{d_{ol}}\left[ 1+e^{(c_{2}-2n\pi )/c_{1}}
 \right] ,
\label{pm7}
\end{equation}
and
\begin{equation}
\zeta _{n}=\frac{u_{ps}}{c_{1}d_{ol}}e^{(c_{2}-2n\pi )/c_{1}}.
\label{pm8}
\end{equation}
Replacing $\theta_n$ in Eq. (\ref{pm2}), $\Delta\alpha_n=(\theta_n-\beta)d_{os}/d_{ls}$. Putting this expression in Eq. (\ref{pm6}), we have
\begin{equation}
\theta _{n}=\theta ^{0}_{n}-\frac{\zeta _{n}d_{os}}{d_{ls}}(\theta _{n}-\beta );
\label{pm10}
\end{equation}
then, using that $0<\zeta_n d_{os}/d_{ls}<1$ and keeping only the first-order term in $\zeta_n d_{os}/d_{ls}$, the angular positions of the images finally take the form
\begin{equation}
\theta _{n}=\theta ^{0}_{n}+\frac {\zeta _{n}d_{os}}{d_{ls}}(\beta -\theta ^{0}_{n}).
\label{pm14}
\end{equation}
The angular positions of the other set of the relativistic images are obtained analogously, and are given by the expression
\begin{equation}
\theta _{n}=-\theta ^{0}_{n}+\frac {\zeta _{n}d_{os}}{d_{ls}}(\beta +\theta ^{0}_{n}).
\label{pm15}
\end{equation}
                                                                                                                                 
The magnification of the $n$-th relativistic image is obtained from Eq. (\ref{magnif1}), with $\theta $ replaced by $\theta _n$; using  Eq. (\ref{pm14}) and that the angles are small, we have
\begin{equation}
\mu _{n}=\frac{1}{\beta}\left[ \theta ^{0}_{n}+
\frac {\zeta _{n}d_{os}}{d_{ls}}(\beta - \theta ^{0}_{n})\right]
\frac {\zeta _{n}d_{os}}{d_{ls}},
\label{pm18}
\end{equation}
and performing a first order Taylor expansion in $\zeta_n d_{os}/d_{ls}$, we finally obtain
\begin{equation}
\mu _{n}=\frac{1}{\beta}\frac{\theta ^{0}_{n}\zeta _{n}d_{os}}{d_{ls}}.
\label{pm19}
\end{equation}
For the other set of relativistic images, the expression of the  $n$-th magnification is also given by Eq. (\ref{pm19}). From Eq. (\ref{pm7}) and (\ref{pm8}), it can be seen that the magnifications decrease exponentially with $n$, which means that the first relativistic image is the brightest one. The magnifications are very faint because they are proportional to $(u_{ps}/d_{ol})^2$, which is a very small factor, unless the lens and the source are highly aligned ($\beta$ close to zero).

\begin{figure}[t!]
\begin{center}
    \includegraphics[scale=0.90,clip=true,angle=0]{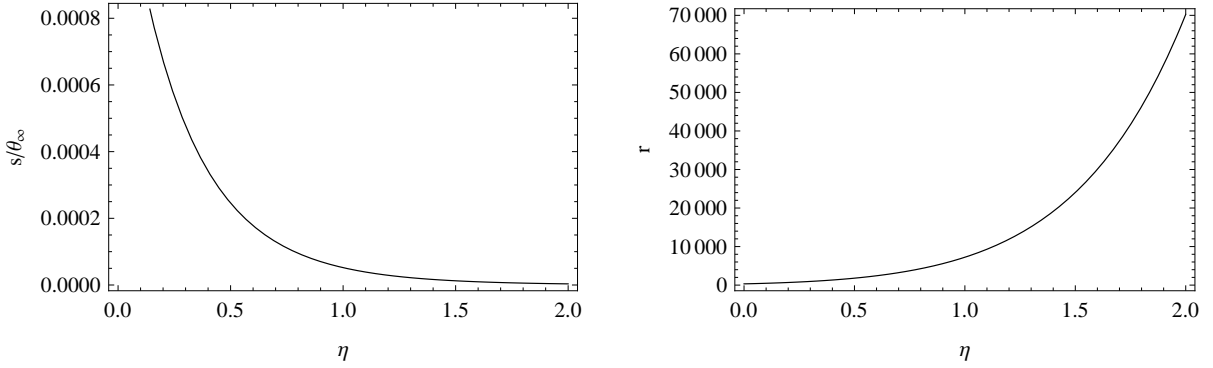}
    \caption{Strong deflection limit observables $s/\theta_\infty$ and $r$ as functions of the parameter $\eta$.}
    \label{obs}
\end{center}
\end{figure}

To relate the results obtained analytically with observations, following \cite{boz} we can define the observables:
\begin{equation}
\theta_{\infty}=\frac{u_{ps}}{d_{ol}},
\label{ob1}
\end{equation}
\begin{equation}
s=\theta_1-\theta_{\infty},
\label{ob2}
\end{equation}
and
\begin{equation}
r=\frac{\mu_1}{\sum^{\infty}_{n=2}\mu_n}.
\label{ob3}
\end{equation}
The quantity $s$ is the angular separation between the position of the first relativistic image and the limiting value of the others $\theta_{\infty}$, and $r$ is the quotient between the flux of the first image and the flux coming from all the other images. As shown in \cite{boz}, for high alignment, these expressions can be reduced to the form
\begin{equation}
s=\theta_{\infty}e^{(c_2-2\pi)/c_1},
\label{ob4}
\end{equation}
and
\begin{equation}
r=e^{2\pi/c_1}.
\label{ob5}
\end{equation}
The values of these observables, as functions of $\eta$ are shown in Fig. \ref{obs}. From the plots, we see that the relative angular separation of the images decreases with $\eta $, and the first image grows in intensity with respect to the others as $\eta $ increases. By measuring $\theta_{\infty}$, $s$ and $r$ and inverting Eqs. (\ref{ob4}) and (\ref{ob5}), the coefficients of the strong deflection limit $c_1$ and $c_2$ can be calculated, and they can be compared with the values predicted by the theoretical models to identify the nature of the black hole acting as a lens.

\section{Discussion}

In this article, we have studied the gravitational lensing by braneworld black holes with no matter and mass. These black holes are characterized by two parameters: $h$ and $\eta $ (or $\kappa $), which come from the tidal effects from the bulk on the brane. The effective energy-momentum tensor is rather peculiar, because it does not satisfy the usual energy conditions in the region outside the horizon. We have analyzed in detail the image production for high alignment, by using the weak and the strong deflection limits.

The value of the parameter $h$ that appears in the metric functions gives the size of the object, i.e. $r_h=h$, so by measuring $\theta _\infty $ and the observer-lens distance one can obtain $h$. As $\theta _\infty $ should be large enough to be resolved by the available or future instruments, $h$ should be large for distant black holes and can have smaller values if these objects exist nearby. The parameter $\eta $ is obtained from $c_1$ and $c_2$; although large values of $\eta$ will be very difficult to find, because in this case the first relativistic image is very close to the others to be separated and it is also too bright compared to them.  

It is interesting to compare the results obtained in the present work, with those corresponding to the spherically symmetric black hole solutions in General Relativity \cite{eiroto,boz}, i.e. the Schwarzschild and the Reissner--Norsdtr\"om geometries. For Schwarzschild black holes $c_1^{Sch}=1$ and $c_2^{Sch}=\ln[216(7-4\sqrt{3})] - \pi \approx -0.4002$, in our case there is a small value of $\eta $ for which $c_1=1$, but $c_2<-0.6$ for any value of $\eta $. In the case of Reissner--Norsdtr\"om metric, $c_1^{RN}\ge 1$ and grows with the absolute value of the electromagnetic charge $Q$, then for small $\eta $ and $Q$, the coefficients $c_1$ and $c_1^{RN}$ can take the same value; but $c_2$ is clearly in a smaller range of values than $c_2^{RN}$. Then, if the strong deflection limit coefficients can be obtained from observational data, the massless black hole studied in the present work can be clearly distinguished from the Schwarzschild and the Reissner--Norsdtr\"om solutions. 

We can also compare our results with those previously obtained for two other braneworld black hole spacetimes. In the case of the Myers--Perry geometry, the strong deflection limit coefficients \cite{bwlens2} are $c^{MP}_1=\sqrt{2}/2\approx 0.707$ and $c^{MP}_2=\sqrt{2}\ln (4\sqrt{2})-\pi \approx -0.691$; we can see from Fig. \ref{coef} that the pair of values $\{ c_1$, $c_2 \}$ calculated here, never coincide with $\{ c^{MP}_1$, $c^{MP}_2\}$, for any value of $\eta $. For tidal charge black holes, from Fig. 4 of Ref. \cite{bwlens3} one can see that coefficient $c^{TC}_1$ grows with the tidal charge, while the coefficient $c^{TC}_2$ decreases with it; by comparing with our Fig. \ref{coef}, it is clear that the black holes can be easily distinguished by their strong deflection limit coefficients. 

The observation of the relativistic images is a major goal in astrophysics, since they correspond to a full description of the near horizon region of black holes. New observational facilities, most of them space-based, will be fully operational in the next years, and will be able to measure in the radio and X bands. We can mention three of them, namely, RADIOASTRON, MAXIM and Event Horizon Telescope. The first one is a space-based radio telescope, launched in July 2011, which will be capable of carrying out measurements with high angular resolution, about $1-10 \, \mu \mathrm{as}$ \cite{zakharov,webradio}. The MAXIM project \cite{webmaxim} is a space-based X-ray interferometer with an expected angular resolution of about $0.1 \, \mu \mathrm{as}$. The third one is a project based on very long baseline interferometry, which proposes to combine existing and planned millimeter/submillimeter facilities into a high-sensitivity, high angular resolution telescope \cite{eht}. Several observational aspects of the 
Galactic center black hole, including some strong deflection features, are discussed in the recent review \cite{mmg}. Nevertheless, it seems that subtle effects, like the comparison of different models of black holes corresponding to alternative theories, will need of a second generation of future instruments. If the braneworld model is a suitable description of the Universe and the massless black holes studied here exist, a very large resolution and a high sensitivity will be necessary for the observation of the effects discussed in this work.

\section*{Acknowledgments}

This work has been supported by Universidad de Buenos Aires and CONICET. We want to thank an anonymous referee for careful reading of the manuscript and for helpful comments.


\begin{thebibliography}{99}

\bibitem{bwrev} D. Langlois, Prog. Theor. Phys. Suppl. \textbf{148}, 181
(2002); P. Brax and C. van de Bruck, Class. Quantum Grav. \textbf{20}, R201
(2003); R. Maartens and K. Koyama, Living Rev. Relativity \textbf{13}, 5 (2010).

\bibitem{kanti} P. Kanti, Int. J. Mod. Phys. A \textbf{19}, 4899 (2004).

\bibitem{rsII} L. Randall and R. Sundrum, Phys. Rev. Lett. \textbf{83}, 3370
(1999); L. Randall and R. Sundrum, Phys. Rev. Lett. \textbf{83}, 4690 (1999).

\bibitem{cgl} R. Guedens, D. Clancy, and A.R. Liddle, Phys. Rev. D \textbf{66}, 043513 (2002); R. Guedens, D. Clancy, and A.R. Liddle, Phys. Rev. D \textbf{66}, 083509 (2002); D. Clancy, R. Guedens, and A.R. Liddle, Phys. Rev. D \textbf{68}, 023507 (2003).

\bibitem{majumdar} A.S. Majumdar, Phys. Rev. Lett. \textbf{90}, 031303 (2003).

\bibitem{dmpr} N. Dadhich, R. Maartens, P. Papadopoulos, and V. Rezania, Phys. Lett. B \textbf{487}, 1 (2000).

\bibitem{bronnikov} K.A. Bronnikov, V.N. Melnikov, and H. Dehnen, Phys. Rev. D
\textbf{68}, 024025 (2003).

\bibitem{darwin} C. Darwin, Proc. Roy. Soc London A \textbf{249}, 180 (1959).

\bibitem{otros} J.-P. Luminet, Astron. Astrophys. \textbf{75}, 228 (1979);
H.C. Ohanian, Am. J. Phys. \textbf{55}, 428 (1987); R.J. Nemiroff, Am. J.
Phys. \textbf{61}, 619 (1993); V. Bozza, S. Capozziello, G. Iovane, and G.
Scarpetta, Gen. Relativ. Gravit. \textbf{33}, 1535 (2001).

\bibitem{eiroto} E.F. Eiroa, G.E. Romero, and D.F. Torres, Phys. Rev. D
\textbf{66}, 024010 (2002).

\bibitem{boz} V. Bozza, Phys. Rev. D \textbf{66}, 103001 (2002).

\bibitem{numerical} K.S. Virbhadra, and G.F.R. Ellis, Phys. Rev. D \textbf{62}, 084003 (2000); K.S. Virbhadra, and C.R. Keeton, Phys. Rev. D \textbf{77}, 124014 (2008); K.S. Virbhadra, Phys. Rev. D \textbf{79}, 083004 (2009).

\bibitem{alternative}  A. Bhadra, Phys. Rev. D  \textbf{67}, 103009 (2003); E.F. Eiroa, Phys. Rev. D \textbf{73}, 043002 (2006); K. Sarkar, and A. Bhadra, Class. Quantum Grav. \textbf{23}, 6101 (2006); N. Mukherjee, and A.S. Majumdar, Gen. Relativ. Gravit. \textbf{39}, 583 (2007); G. N. Gyulchev, and S. S. Yazadjiev, Phys. Rev. D \textbf{75}, 023006 (2007); S. Chen and J. Jing, Phys. Rev. D \textbf{80}, 024036 (2009); E.F. Eiroa and C.M. Sendra, Class. Quantum Grav. \textbf{28},  085008 (2011). 

\bibitem{nakedsing1}  K.S. Virbhadra, D. Narasimha, and S.M. Chitre, Astron. Astrophys. \textbf{337}, 1 (1998).

\bibitem{nakedsing2} K.S. Virbhadra, and G.F.R. Ellis, Phys. Rev. D \textbf{65}, 103004 (2002).

\bibitem{bwlens1} V. Frolov, M. Snajdr, and D. Stojkovic,  Phys. Rev. D \textbf{68}, 044002 (2003).

\bibitem{bwlens2} E.F. Eiroa, Phys. Rev. D \textbf{71}, 083010 (2005); E.F. Eiroa, Braz. J. Phys. \textbf{35}, 1113 (2005). 

\bibitem{bwlens3} R. Whisker, Phys. Rev. D \textbf{71}, 064004 (2005).

\bibitem{bwlens4}  A.S. Majumdar, and N. Mukherjee, Int. J. Mod. Phys. D \textbf{14}, 1095 (2005); C. R. Keeton, and A. O. Petters, Phys. Rev. D \textbf{73}, 104032 (2006); S. Pal and S. Kar, Class. Quantum Grav. \textbf{25}, 045003 (2008); A.Y. Bin-Nun, Phys. Rev. D \textbf{81}, 123011 (2010);  A.Y. Bin-Nun, Phys. Rev. D \textbf{82}, 064009 (2010).

\bibitem{bozza1} V.Bozza, Phys. Rev. D \textbf{67}, 103006 (2003); V. Bozza, F. De Luca, G. Scarpetta, and M. Sereno, Phys. Rev. D \textbf{72}, 083003 (2005); V. Bozza, F. De Luca, and G. Scarpetta, Phys. Rev. D \textbf{74}, 063001 (2006).

\bibitem{vazquez} S. V\'azquez and E. Esteban, Nuovo Cim. \textbf{119B}, 489 (2004).

\bibitem{bozza2} V. Bozza, G. Scarpetta, Phys. Rev. D \textbf{76}, 083008 (2007).

\bibitem{kraniotis} G.V. Kraniotis, Class. Quantum Grav. \textbf{28}, 085021 (2011).

\bibitem{falcke} H. Falcke, F. Melia, and E. Agol, Astrophys. J. \textbf{528}, L13 (2000).

\bibitem{shadow} A. de Vries, Class. Quant. Grav. \textbf{17}, 123 (2000); R. Takahashi, Astrophys. J. \textbf{611}, 996 (2004); K. Hioki and U. Miyamoto, Phys. Rev. D \textbf{78}, 044007 (2008); C. Bambi, and K. Freese, Phys. Rev. D \textbf{79}, 043002 (2009); K. Hioki and K.I. Maeda, Phys. Rev. D \textbf{80}, 024042 (2009); L. Amarilla, E.F. Eiroa, and G. Giribet, Phys. Rev. D \textbf{81}, 124045 (2010).

\bibitem{zakharov} A.F. Zakharov, A.A. Nucita, F. DePaolis, and G. Ingrosso, New Astron. \textbf{10}, 479 (2005); A.F. Zakharov, F. De Paolis, G. Ingrosso, and A.A. Nucita, Astron. Astrophys. \textbf{442}, 795 (2005); F. De Paolis, G. Ingrosso, A.A. Nucita, A. Qadir, and A.F. Zakharov, Gen. Relativ. Gravit. \textbf{43}, 977 (2011).

\bibitem{aliev} A.N. Aliev and A.E. G\"umr\"uk\c{c}\"uoglu, Phys. Rev. D \textbf{71}, 104027 (2005).

\bibitem{shbw} A.N. Aliev and P. Talazan, Phys. Rev. D \textbf{80}, 044023 (2009); J. Schee and Z. Stuchlik, Int. Jour. Mod. Phys. D \textbf{18}, 983 (2009); L. Amarilla and E.F. Eiroa, Phys. Rev. D \textbf{85}, 064019 (2012).

\bibitem{bozzareview} V. Bozza, Gen. Relativ. Gravit. \textbf{42}, 2269 (2010).

\bibitem{weinberg}  S. Weinberg, \textit{Gravitation and Cosmology: Principles and Applications of the General Theory of Relativity} (Wiley, New York, 1972).

\bibitem{bozzale} V. Bozza, Phys. Rev. D \textbf{78}, 103005 (2008).

\bibitem{webradio} http://www.asc.rssi.ru/radioastron

\bibitem{webmaxim} http://maxim.gsfc.nasa.gov

\bibitem{eht} http://eventhorizontelescope.org

\bibitem{mmg} M.R. Morris, L. Meyer, and A.M. Ghez, Res. Astron. Astrophys. \textbf{12}, 995 (2012).

\end{thebibliography}
\end{document}